# Microcanonical Foundation for Systems with Power-Law Distributions

Sumiyoshi Abe[1] and A. K. Rajagopal[2]

[1]*College of Science and Technology, Nihon University,*

*Funabashi, Chiba 274-8501, Japan*

[2]*Naval Research Laboratory, Washington, D. C. 20375-5320*

Starting from microcanonical basis with the principle of equal *a priori* probability, it is found that, besides ordinary Boltzmann-Gibbs theory with the exponential distribution, a theory describing systems with power-law distributions can also be derived.

Systems exhibiting (quasi) power-law behavior in their probability distributions are quite ubiquitous in nature [1,2]. It is a challenging problem to understand the properties of such systems based on principles of statistical mechanics, because these systems can hardly be described by ordinary Boltzmann-Gibbs canonical ensemble theory, whose distribution is of the exponential form. In this Letter, we show that it is actually possible to derive non-Boltzmann-Gibbs theory from the microcanonical basis with the principle of equal *a priori* probability. The resulting distribution is found to exhibit the required power law.

We first recall the Gibbs theorem [3,4], which states that a small subsystem of a microcanonical ensemble with large degrees of freedom is *uniquely* characterized by the standard exponential distribution. Historically this theorem has repeatedly been proved in various ways, such as the method of counting, the method of steepest descents, and the application of the central limit theorem. Starting from the microcanonical basis with principle of equal *a priori* probability, one obtains the exponential distribution for the canonical ensemble. Therefore, if the Gibbs theorem were universal, then any of the equilibrium theories other than Boltzmann-Gibbs theory could not exist and consequently power-law distributions would be excluded, as long as microcanonical ensemble theory is the basis. Here, we show that a route to canonical ensemble theory for systems with power-law distributions is actually allowed within the microcanonical ensemble theory.

To exhibit this route, we begin with the standard discussion [5] about the Gibbs theorem by considering a classical system $s$ and take its $N$ replicas $s_1, s_2, \text{L}, s_N$. The collection $\text{S} = \{s_\alpha\}_{\alpha=1,2,\text{L},N}$ is referred to as a supersystem. Let $A_\alpha$ be a physical quantity associated with the system $s_\alpha$. It is a statistical random variable and its value is denoted by $a(m_\alpha)$, where $m_\alpha$ labels the allowed configurations of $s_\alpha$. The quantity of physical interest is the average of $\{A_\alpha\}_{\alpha=1,2,\text{L},N}$ over the supersystem: $(1/N)\sum_{\alpha=1}^{N} A_\alpha$. According to microcanonical ensemble theory, the probabilities of finding $\text{S}$ in the configurations in which the values of the average quantity lies around a certain value $\bar{a}$, i.e.,

$$\left| \frac{1}{N} \sum_{\alpha=1}^{N} a(m_\alpha) - \bar{a} \right| < \varepsilon, \tag{1}$$

are all equal. $\varepsilon$ is assumed to be

$$\varepsilon \sim O(N^{-1-\delta}), \tag{2}$$

where $\delta > 0$. The equiprobability $\mathrm{P}\,(m_1, m_2, \mathrm{L}\,, m_N)$ associated with this condition is

$$\mathrm{P}\,(m_1, m_2, \mathrm{L}\,, m_N) \propto \theta\!\left( \varepsilon - \left| \frac{1}{N} \sum_{\alpha=1}^{N} a(m_\alpha) - \bar{a} \right| \right), \tag{3}$$

where $\theta(x)$ stands for the Heaviside unit-step function. To shift from microcanonical ensemble theory to canonical ensemble theory, we fix the objective system and eliminate the other ones. The probability of finding the objective system, say $s_1$, in the configuration $m_1 = m$ is given by

$$p(m) = \sum_{m_2, \mathrm{L}\,, m_N} \mathrm{P}\,(m, m_2, \mathrm{L}\,, m_N), \tag{4}$$

which characterizes the canonical ensemble.

To prove the Gibbs theorem, the following integral representation of the step function is employed:

$$\theta(x) = \int_{\beta-i\infty}^{\beta+i\infty} d\phi\, \frac{e^{\phi x}}{2\pi i \phi}, \tag{5}$$

where $\beta$ is an arbitrary positive constant. Then, in the large-$N$ limit, one applies the method of steepest descents to evaluate the integration over $\phi$. Contextually, it is clear that the exponential distribution in Boltzmann-Gibbs canonical ensemble theory has its origin in this integral representation of $\theta(x)$ using the exponential function.

To examine the possibility of obtaining the power-law-type distribution, we consider the "$q$-exponential function", which is given by

$$e_q(x) \equiv \left[1+(1-q)x\right]^{1/(1-q)}, \tag{6}$$

where $q$ is a real parameter satisfying the condition

$$q > 1. \tag{7}$$

In the limit $q \to 1$, $e_q(x)$ converges to the ordinary exponential function. An important point is that even if the exponential function in the integrand in eq. (5) is replaced by the $q$-exponential function, still the equality

$$\theta(x) = \int_{\beta-i\infty}^{\beta+i\infty} d\phi \, \frac{e_q(\phi x)}{2\pi i \phi}, \tag{8}$$

as long as $\beta$ is taken to satisfy

$$1-(q-1)\beta x_{\max} > 0. \tag{9}$$

The simplest way to ascertain this fact may be to use the following integral representation:

$$e_q(\phi x) = \frac{1}{\Gamma\left(\frac{1}{q-1}\right)} \int_0^\infty dt\, t^{\frac{1}{q-1}-1} \exp\{-[1+(1-q)\phi x]t\}, \tag{10}$$

where $\Gamma(s)$ is the gamma function. The condition in eq. (9) will be discussed later in the context of the steepest descent approximation.

Equation (8) may be understood from the point of view that the step function is of discrete topology and therefore can remain invariant under continuous deformation of the exponential function in the integrand in eq. (5). In this respect, it is clear that the use of the $q$-exponential function is nothing but one particular choice of deformation.

Using eq. (8), we have, *in the leading order in* $N$,

$$\theta\!\left(\varepsilon - \left[\frac{1}{N}\sum_{\alpha=1}^N a(m_\alpha) - \bar{a}\right]\right) = \int_{\beta-i\infty}^{\beta+i\infty} \frac{d\phi}{2\pi i\phi} e_q(\phi\varepsilon) \prod_{\alpha=1}^N e_q\!\left(-\phi\frac{1}{N}[a(m_\alpha) - \bar{a}]\right). \tag{11}$$

To obtain this equation, first we factorize the $q$-exponential function as $e_q(\phi[\varepsilon - M]) \approx e_q(\phi\varepsilon)e_q(-\phi M)$ with $M \equiv (1/N)\sum_{\alpha=1}^N a(m_\alpha) - \bar{a}$ and then further factorize $e_q(-\phi M)$ into the product over the supersystem. Since $|M| < \varepsilon \sim O(N^{-1-\delta})$ due to eqs. (1) and (2), this manipulation will be justified in the subsequent discussion of the method of steepest descents in the large-$N$ limit.

Noting that $\theta(\varepsilon - |M|) = \theta(\varepsilon - M) - \theta(-\varepsilon - M)$ and changing the integration variable as $\phi \to N\phi$, we obtain

$$\theta\!\left(\varepsilon - \left|\frac{1}{N}\sum_{\alpha=1}^N a(m_\alpha) - \bar{a}\right|\right) = \int_{\beta^*-i\infty}^{\beta^*+i\infty} d\phi\, \frac{\sinh_q(N\phi\varepsilon)}{\pi i\phi} \prod_{\alpha=1}^N e_q(-\phi[a(m_\alpha) - \bar{a}]), \tag{12}$$

where $\sinh_q(x) \equiv [e_q(x) - e_q(-x)]/2$ and

$$\beta^* = \frac{\beta}{N}. \tag{13}$$

Here, we examine the condition in eq. (9). In the present context, it is written as

$$1-(q-1)\beta^* N \left| \pm \varepsilon - M \right|_{\max} > 0. \tag{14}$$

The rectangular distribution function we are considering has a very narrow support. In fact, $\left| \pm \varepsilon - M \right|_{\max}$ is a quantity of $O(N^{-1-\delta})$ with $\delta > 0$. Therefore, in the large-$N$ limit, $\beta^*$ can be an arbitrary positive constant.

Thus, working out to the leading order in $N$, we can express the probability as follows:

$$p_q(m) = \sum_{m_2, \ldots, m_N} \text{P}\left(m, m_2, \ldots, m_N\right)$$

$$= \frac{1}{W} \int_{\beta^*-i\infty}^{\beta^*+i\infty} d\phi \, \frac{\sinh_q(N\phi\varepsilon)}{\pi i \phi} e_q\left(-\phi\left[a(m)-\bar{a}\right]\right)$$

$$\times \sum_{m_2, \ldots, m_N} \prod_{\alpha=2}^{N} e_q\left(-\phi\left[a(m_\alpha)-\bar{a}\right]\right)$$

$$= \frac{1}{W} \int_{\beta^*-i\infty}^{\beta^*+i\infty} d\phi \, \frac{\sinh_q(N\phi\varepsilon)}{\pi i \phi} \frac{e_q\left(-\phi\left[a(m)-\bar{a}\right]\right)}{\tilde{Z}_q(\phi)} \exp\left[N \ln \tilde{Z}_q(\phi)\right], \tag{15}$$

where $W$ is the number of possible configurations satisfying eq. (1) and

$$\tilde{Z}_q(\phi) = \sum_m e_q\left(-\phi\left[a(m)-\bar{a}\right]\right). \tag{16}$$

Clearly,

$$W = \int_{\beta^*-i\infty}^{\beta^*+i\infty} d\phi \, \frac{\sinh_q(N\phi\varepsilon)}{\pi i \phi} \exp\left[N \ln \tilde{Z}_q(\phi)\right], \tag{17}$$

since $p_q(m)$ is normalized. Now, using the real part $\beta^*$ of $\phi$, the steepest descent condition reads

$$\frac{\partial \tilde{Z}_q}{\partial \beta^*} = 0, \tag{18}$$

which leads to

$$p_q(m) = \frac{1}{\tilde{Z}_q(\beta^*)} e_q\{-\beta^*[a(m) - \bar{a}]\}, \tag{19}$$

$$\bar{a} = \sum_m P_q(m) a(m), \tag{20}$$

simultaneously, where $P_q(m)$ is given by

$$P_q(m) = \frac{[p_q(m)]^q}{\sum_m [p_q(m)]^q}. \tag{21}$$

These results follow from eq. (6) and the relation $d e_q(x)/dx = [e_q(x)]^q$. It is evident that in the limit $q \to 1+0$ all the discussions become reduced to the ordinary ones in Boltzmann-Gibbs theory with the familiar canonical distribtuion of the exponential form. On the other hand, the distribution function in eq. (19) is seen to asymptotically exhibit the power-law behavior

$$p_q(m) \sim \frac{1}{[a(m)]^{1/(q-1)}}, \tag{22}$$

as desired.

In the field of thermodynamics of chaotic systems [6], $P_q(m)$ in eq. (21) is referred to as the escort distribution, which is also a probability distribution associated with the original distribution $p_q(m)$. The steepest descent condition yields the fact that the arithmetic mean of $\{A_\alpha\}_{\alpha=1,2,\text{L},N}$ coincides with the generalized expectation value with respect to the escort distribution as in eq. (20).

In conclusion, we have shown that not only ordinary Boltzmann-Gibbs canonical ensemble theory but also a theory for systems with power-law distributions can be obtained from microcanonical basis with the principle of equal *a priori* probability. It is worth pointing out that the structure in eqs. (19)-(21) is the characteristic of nonextensive statistical mechanics [7]. It is known that generalized canonical ensemble theory derived from maximum entropy principle based on the Tsallis entropy [8] with eq. (20) as the constraint gives rise to the distribution in eq. (20). However, it is essential to recall that here we made no initial assumptions on the definition of statistical expectation value and the form of the entropy.

The authors would like to thank Professor R. Balian for drawing their attention to Ref. [5]. S. A. was supported in part by the GAKUJUTSU-SHO Program of College of Science and Technology, Nihon University. A. K. R. acknowledges the support of the U. S. Office of Naval Research.